\documentclass[12pt,a4]{article}
\usepackage{amsmath}
\usepackage{bm}
\usepackage{amsfonts}
\usepackage{amssymb}
\usepackage{graphics}
\usepackage[normal]{caption2}
\usepackage{subfigure}
\usepackage{rotating}
\usepackage{citesort}
\def\be{\begin{equation}}
\def\ee{\end{equation}}
\def\ba{\begin{array}}
\def\ea{\end{array}}
\def\bea{\begin{eqnarray}}
\def\eea{\end{eqnarray}}

\begin{document}
\baselineskip 20pt \setlength\tabcolsep{2.5mm}
\renewcommand\arraystretch{1.5}
\setlength{\abovecaptionskip}{0.1cm}
\setlength{\belowcaptionskip}{0.5cm}
\begin{center} {\large\bf Alpha decay chains study for the recently observed superheavy element $Z=117$ within the Isospin Cluster Model}\\
\vspace*{0.4cm}
{\bf Sushil Kumar}
\footnote{Email:~sushilk17@gmail.com}\\
$^a${\it  Department of Applied Science, Chitkara University, Solan
-174103,(H.P.) India.\\}

\end{center}
The recently observed $\alpha$-decay chains $^{293-294}117$ were produced by the fusion reactions with target $^{249}Bk$ and projectile $^{48}Ca$ at Dubna in Russia. The reported cross-sections for the mentioned reaction are $\sigma=0.5(+1.1,-0.4)$pb and $\sigma$=1.3(+1.5,-0.6)$pb$ at $E^{*}=35MeV$ and $E^{*}=39MeV$, respectively.
The Q-values of $\alpha$-decay and the half-lives $Log_{10}T^{\alpha}_{1/2}$(s) are calculated for the $\alpha$-decay chains of $^{293-294}117$ nuclei, within the framework of Isospin Cluster Model (ICM). In the ICM model the proximity energy is improved by using the isospin dependent radius of parent, daughter and alpha particle. The binding energy $B(A_{i}, Z_{i})$ (i=1,2) of any nucleus of mass number A and atomic number Z was obtained from a phenomenological and more genaralized BW formula given by \cite{samanta02}. The calculated results in ICM are compared with the experimental results and other theoretical Macro-Microscopic(M-M), RMF( with NL3 and SFU Gold forces parameter) model calculations. The estimated values of $\alpha$-decay half-lives are in good agreement with the recent data. The ICM calculation is in favor of the persence of magic number at $N=172$.   

\newpage
\baselineskip 20pt
\section{Introduction}
In the nuclear chart, the superheavy mass region are growing fast due to the availability and advancement in the radioactive nuclear beam technology. The $^{48}Ca$ is most prominent radioactive nuclear beam at present for the synthesis of superheavy elements \cite{yuri06, yuri07}. Flerov \cite{ref1} was the first who suggested in 1969, the use of a highly neutron-rich beam of $^{48}Ca$ for the formation of superheavy elements with neutron rich targets such as $^{244}Pu$, $^{248}Cm$ and $^{252}Cf$. The stability of these superheavy nuclei depends upon the magicity of the proton and neutron number either spherical and/or deformed. The spherical shell closure for the neutron and proton numbers are at 2, 8, 20, 28, 50, and 82 while 126 is only for neutron number. Since a long time, the search about the next spherical shell for the proton number is going on. This question, about the next doubly magic nucleus beyond the $^{208}Pb$ ($Z=82, N=126$) has attracted much attention in the nuclear structure physics for both the theoreticans and experimentalist. Theoretical models predict \cite{cwiok96,rutz97,kruppa00,bender01,bender03} that the next magicity for the proton number should occur at $Z = 114, 120, or 126$ and for the neutron number it should be at $N = 172$ and $184$.\\

These superheavy nuclei undergo spontaneous decay in to successive alpha decay chains before spontaneous fission. The $\alpha$-decay energy and half-lives of these decay chains help us to understand the nuclear structure of the parents as well as of daughter nuclei and hence gives the information about the stability of peninsula. In this paper, recently observed two isotopes of the element Z=117 with mass number 293 and 294 are studied for the alpha decay characteristics. These two isotopes $^{293}117$ and $^{294}117$ were produced in the fusion reaction between $^{48}Ca$ projectile and radioactive target $^{249}Bk$ nuclei \cite{ogane10}. The resulting excitation energy of the compound nucleus $^{297}117$ is reported $E^{*}=39 MeV$ and  $E^{*}=35 MeV$ respectively, which favoring the 4n and 3n evaporation channel. The Isospin Cluster Model (ICM) calculations for the $^{293,294}117$ alpha decay chains are compared with the experimental results \cite{ogane10}, the macroscopic- micrscopic(MM)\cite{sobi10}, RMF based (using NL3 and SFU Gold forces parameters)\cite{patra11,roy11} model calculations.
In our earlier work for the $\alpha$-decay chains calculation \cite{sushil02,sushil03,sushil09}, preformed cluster model \cite{malik89} was used and it has been observed that calculated results could be improved using the isospin physics in the model. So, the earlier used model is modified with the same idea and hence obtained very nice results from the present ICM calculations.\\

The $^{293,294}117$ alpha decay chains calculation are based on the Isospin Cluster Model (ICM), which is described briefly in Section 2 and the results of calculation are presented in Section 3. A discussion and summary of results is given in Section 4.

 \par
\section {The Isospin Cluster Model}
The ICM model uses the dynamical collective coordinates of mass (and charge) asymmetry,
$\eta={{(A_1-A_2)}/{(A_1+A_2)}}$ and $\eta _Z={{(Z_1-Z_2)}/{(Z_1+Z_2)}}$,
first introduced in the QMFT \cite{maruhn74,gupta75}, which are in addition to the usual coordinates of relative separation R and deformations $\beta_{2i}$ ($i=1,2$) of two fragments. Then, in the standard approximation of decoupled R and $\eta $ motions, the decay constant $\lambda$ or the decay half-life $T_{1/2}$ is defined as
\be
\lambda ={{{ln 2}\over {T_{1/2}}}}=P_0\nu _0 P.
\label{eq:1}
\ee
Here $P_0$ is the cluster (and daughter) preformation probability and P the
barrier penetrability which refer, respectively, to the $\eta$ and R motions.
The $\nu _0$ is the barrier assault frequency. The $P_0$ are the solutions of
the stationary Schr\"odinger equation in $\eta$,
\be
\{ -{{\hbar^2}\over {2\sqrt B_{\eta \eta}}}{\partial \over {\partial
\eta}}{1\over {\sqrt B_{\eta \eta}}}{\partial\over {\partial \eta
}}+V_R(\eta )\} \psi ^{({\nu})}(\eta ) = E^{({\nu})} \psi ^{({\nu})}(\eta ),
\label{eq:2}
\ee
which on proper normalization are given as
\be
P_0={\sqrt {B_{\eta \eta}}}\mid \psi ^{({0})}(\eta (A_i))\mid ^2\left ({2/A}\right ),
\label{eq:3}
\ee
with i=1 or 2 and $\nu$=0,1,2,3....
Eq. (\ref{eq:2}) is solved at a fixed $R=R_a=C_t(=C_1+C_2)$.
$C_i$'s taken from the Myers and \'{S}wiatecki \cite{myers20} droplet model.
The matter radius $C_i$ is calculated as
\begin{equation}
C_i={c_i + {{{N_i}\over{A_i}} t_i}}    (i=1,2),
\label{eq:4}
\end{equation}
where $c_i$ denotes the half-density radii of the charge distribution and
$t_i$ is the neutron skin of the nucleus. For the $t_i$ calculation Myers and \'{S}wiatecki used two parameter Fermi function values given in Ref.\cite{jager} and remaining cases were handled with the help of parameterization of charge distribution decribed below.
The nuclear charge radius (denoted as $R_{00}$ in Ref.\cite{nerlo}is given by the relation:
\be
\ba{l}
R_{00i}=\sqrt{5 \over 3}\left\langle r^2\right\rangle^{1/2}\\
=1.240A^{1\over 3}_{i} \left\{1+\frac{1.646}{A_i}-0.191\times\left(\frac{A_i-2Z_i}{A_i}\right)\right\}fm (i=1, 2),
\ea
\label{eq:5}
\ee
where $<r^2>$ repersents the mean square nuclear charge
radius. According to Ref.\cite{nerlo}, Eq.(5) was valid for the even-even nuclei with
$8\leq Z > 38$ only. For nuclei with Z $\geq$ 38, the above equation was
modified by Pomorski et al.,\cite{nerlo} as
\be
R_{00i}=1.256{A_i}^{1\over 3}
\left\{1-0.202\left(\frac{A_i-2Z_i}{A_i}\right)\right\}fm (i=1, 2),
\label{eq:6}
\ee
These expressions give good estimate of the measured mean square nuclear charge
radius$<r^2>$. In the present model, author have used only Eq.(5). The half-density
radius, $c_i$ was obtained from the relation:
\be
c_i=R_{00i}\left(1-{\frac{7}{2}\frac{b^2}{R^{2}_{00i}}}-{\frac{49}{8}\frac{b^4}
{R^{4}_{00i}}}+....\right),(i=1,2)
\label{eq:7}
\ee
Using the Droplet model \cite{myers69}, neutron skin $t_i$, reads as
\be
t_i=\frac{3}{2}r_0
\left[\frac{JI_i-\frac{1}{12}c_1Z_iA^{-1/3}_i}{Q+\frac{9}{4}JA^{-1/3}_i}\right], (i=1,2)
\label{eq:8}
\ee
Here $r_0$ is 1.14 fm. the value of nuclear symmetric energy coefficient J=32.65 MeV and $c_1=3e^2/5r_0= 0.757 895$MeV. The neutron skin stiffness coefficient Q was taken to be 35.4MeV.\\

The fragmentation potential $V_R(\eta )$ (at $R=R_a=C_t(=C_1+C_2)$)in (\ref{eq:2}) is
calculated simply as the sum of the Coulomb interaction, the
nuclear proximity potential \cite{blocki77}with new isospin dependent radii and the ground state binding energies of two nuclei,
\be V(C_t, \eta) =- \sum_{i=1}^{2} B(A_{i}, Z_{i})+
\frac{Z_{1} Z_{2} e^{2}}{C_t} + V_{P},
\label{eq:9}
\ee
The proximity potential between two nuclei is defined as
\begin{equation}
V_{p}=4\pi\overline C\gamma b\Phi (\xi )%
\label{eq:10}
\end{equation}
here $\gamma$ is the nuclear surface tension coefficient, $\overline C$ determines the distance between two points of the surfaces, evaluated at the point of closest approach  using eq.(4) and $\Phi (\xi )$ is the universal function, since it depends only on
the distance between two nuclei, and is given as
\begin{equation}
\Phi (\xi )=\left \{
\ba{l}
-0.5 (\xi-2.54)^{2}- 0.0852(\xi-2.54)^{3},\\
for \quad\xi\leq 1.2511 \\
-3.437 exp(-\xi/0.75).\\
for \quad\xi\geq 1.2511
\ea
\right.
\label{eq:11}
\end{equation}
Here, $\xi$= s/b, i.e s in units of b, with the separation distance
s=$R-C_{1}-C_{2}$. b is the diffuseness of the nuclear surface, given by
\begin{equation}
b=\left[\pi /2\sqrt{3} \ln 9\right]_{t_{10-90}}%
\label{eq:12}
\end{equation}
where $t_{10-90}$ is the thickness of the surface
in which the density profile changes from 90$\%$ to 10$\%$.
The $\gamma$ is the specific nuclear surface tension, given by
\begin{equation}
\gamma =0.9517\left[1-1.7826\left(\frac{N-Z}{A} \right)^{2}%
\right] MeV fm^{-2}.%
\label{eq:13}
\end{equation}
In recent years many more microscopic potentials are available that takes care various aspects such as overestimation of fusion barrier in original proximity potential, isospin effects. A comparison is also available between all models \cite{puri10}. The binding energy $B(A_{i}, Z_{i})$ (i=1,2) of any nucleus of mass number A and atomic number Z was obtained from a phenomenological search and was given by a more genaralized BW formula \cite{samanta02} used and found good in agreement with experimental results by many others for drp-line to superheavy nuclei \cite{basu04}. Thus, shell effects are also contained in our calculations in addition to the isospin effects for all the normal to neutron/proton rich nuclei. The momentum dependent potentials and symmtry energy potential which are found to have drastic effect at higher densities will not affect decay studies, since these  happens at lower tale of the density \cite{sood,kumar}. Here in Eq. (4), the Coulomb and proximity potentials are for spherical nuclei, and charges $Z_1$ and $Z_2$ in (4) are fixed by minimizing the potential in $\eta_Z$ coordinate. The mass parameters $B_{\eta \eta}(\eta )$, representing the kinetic energy part in Eq. (\ref{eq:2}), are the classical hydrodynamical masses of Kr\"oger and Scheid \cite{kroeger80}.\\

\begin{figure}[!t]
\centering
\vskip -1cm
\includegraphics[width=12cm]{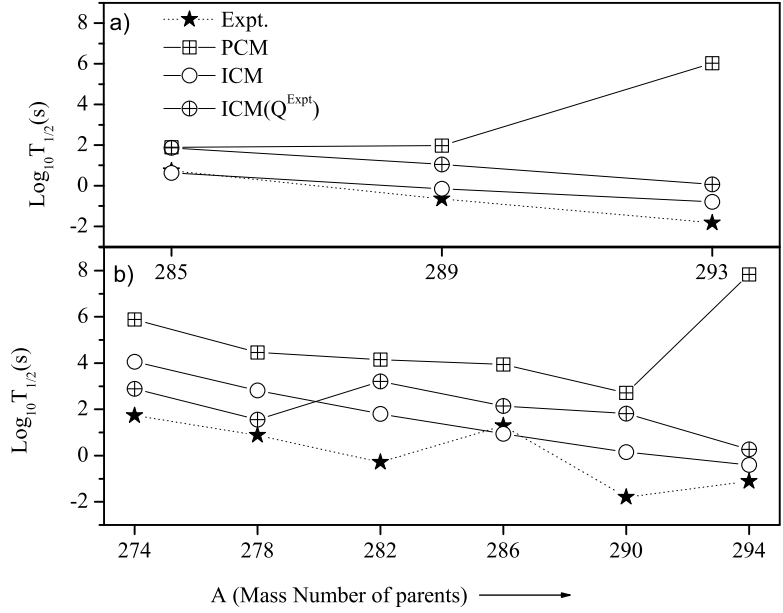}
\caption{(Color online) (a) $log_{10}T^{\alpha}_{1/2}(s)$ for $\alpha$-decay series of $^{293}117$ nucleus.(b) $log_{10}T^{\alpha}_{1/2}(s)$ for $\alpha$-decay series of $^{294}117$ nucleus, calculated on the basis of the Isospin Cluster Model (ICM), compared with the experimental results, preformed cluster model (PCM) and ICM as a function of $Q^{Expt.}_{\alpha}$ calculations, plotted as a function of parent mass number, in these figures.}
\label{fig1}
\end{figure}
The WKB tunnelling probability, calculated is  $P=P_i P_b$ with
\be
P_i=exp[-{2\over \hbar}{{\int }_{R_a}^{R_i}\{ 2\mu [V(R)-V(R_i)]\}
^{1/2} dR}]
\label{eq:14}
\ee
\be
P_b=exp[-{2\over \hbar}{{\int }_{R_i}^{R_b}\{ 2\mu [V(R)-Q]\} ^{1/2} dR}].
\label{eq:15}
\ee
These integrals are solved analytically \cite{malik89} for $R_b$, the second
turning point, defined by $V(R_b)=Q$-value for the ground-state decay.
\begin{figure}[!t]
\centering
\vskip -1cm
\includegraphics[width=12cm]{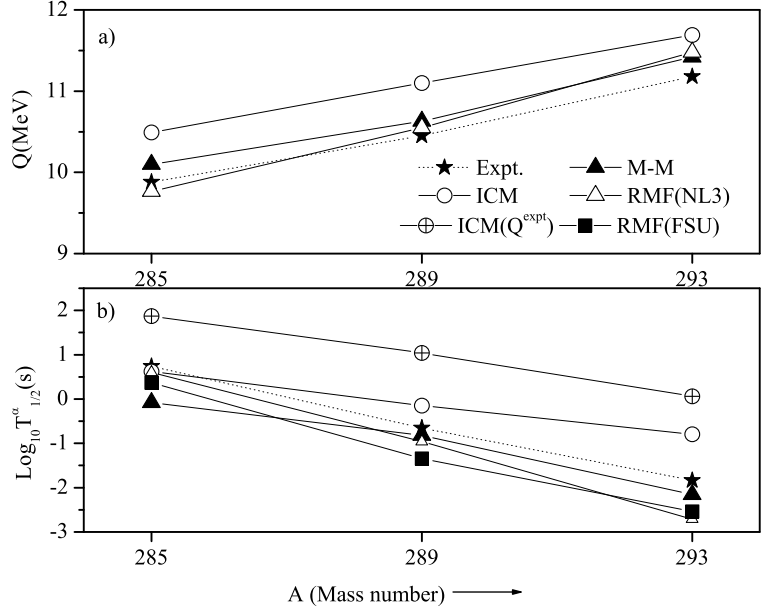}
\caption{(Color online) (a) The $Q^{\alpha}$-energy and (b) the half-life times $log_{10}T^{\alpha}_{1/2}(s)$ for $\alpha$-decay series of $^{293}117$ nucleus, calculated on the basis of Isospin Cluster Model (ICM), compared with the experimental results \cite{ogane10}, Macro-Microscopic \cite{sobi10}, RMF(NL3 parameter) \cite{patra11}, RMF( FSU Gold forces parameter) \cite{roy11} and ICM as a function of $Q^{Expt.}_{\alpha}$ calculations, plotted as a function of parent mass number, in these figures.}
\label{fig2}
\end{figure}

The assault frequency $\nu _0$ in (\ref{eq:1}) is given simply as
\be
\nu _0=(2E_2/\mu )^{1/2}/R_0,
\label{eq:16}
\ee
with $E_2=(A_1/A) Q$, the kinetic energy of the lighter fragment, for the
$Q$-value shared between the two products as inverse of their masses.\\
\begin{figure}[!t]
\centering
\vskip -1cm
\includegraphics[width=12cm]{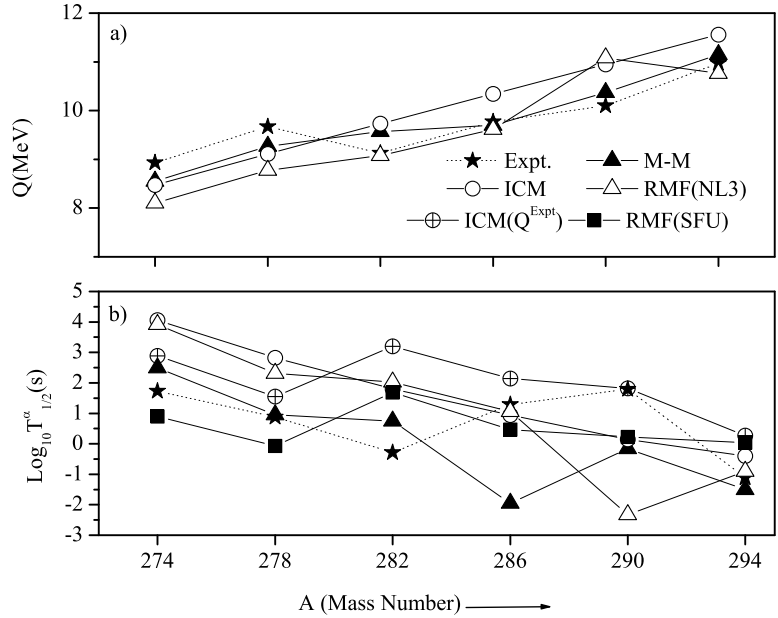}
\caption{(Color online) Same as for Fig. 2, but for the $^{294}117$ nucleus.}
\label{fig3}
\end{figure}

 \par
\section{Calculations and results}
In Fig.1 two $\alpha$-decay chains of the Z=117 isotopes, $^{294}117$ and $^{293}117$ are shown for the alpha decay half-lives and Q-value calculations. A comparision of PCM  and ICM calculations with experimental results are shown. The Isospin Cluster Model calculations are comparable and are in good agreement with experimental results. For both the alpha decay chains ICM calculations also performed as a function of
$Q^{Exp}$. The results ICM($Q^{Exp}$) are resonably well for the $Z=107, 109$ nuclei in comparision to ICM with the others in $^{294}117$ $\alpha$-decay chain. Where $Q^{ICM}$-value is calculated from the BW mass formula suggested by samanta et.al. \cite{samanta02}.
Two $\alpha$-decay chains are calculated within the ICM model framework for the four different Q-values i.e., $Q^{Exp}$,  $Q^{M-M}$,  $Q^{RMF(NL3)}$, and $Q^{ICM}$ to see the effect of the Q-value on the alpha decay half-lives. Which can be seen in the Fig. 2 and 3 clearly. Further the ICM Q-values $Q^{ICM}$ are compared with the experimental ($Q^{Exp}$), $Q^{MM}$,  $Q^{RMF(NL3)}$ values for both the alpha decay chains. In table 1, Fig.2 and 3 the ICM model calculations for the alpha decay half-lives are compared with the experimental \cite{ogane10}, Macro-Microscopic calculations (M-M)\cite{sobi10}, RMF model calculations using NL3 \cite{patra11} and FSU Gold forces \cite{roy11} parameters. In Fig. 2b the calculated ICM half-lives results are comparable and reasonably well with experimental and other theoretical model calculations. The ICM half-lives as a function of $Q^{Expt.}$ are also studied and shown in figure.\\

Similarly, in Fig. 3, a comparision of different Q-values and half-lives are shown for the $^{294}117$. The calculated ICM and ICM function of ($Q^{Expt.}$), half-lives are in good agreement with experimental results except to the $^{282}111$ ($Z=111, N=171$)nucleus, where a noticeable maximum half-life is found for the nucleus indicates the shell effects(N=172). The decay studies shows that half-lives of the alpha decay as well as cluster decay works as a tool in nuclear structure physics to show the persence of shell effects of the parents as well as of daughter nuclei. A higher value of the half-life indicates the presence of shell stabilized parent nucleus, whereas a comparatively low value of half-life tells the same about the daughter and cluster
nuclei \cite{sushil03,sushil09}. The RMF(NL3) calculation shows the same trends and results except to $^{290}115$ ($Z=115, N=175$) nucleus, with the ICM calculations. The other RMF calculation using the SFU Gold parameter also follows the same trends as the ICM($Q^{Expt.}$) results.
 \par
\section{Discussion and summary}
The Isospin Cluster Model (ICM) is used for the calculations of the two $^{293}117$ and $^{294}117$ alpha-decay chains. The calculated alpha decay half-lives are compared with the experimental results and the other theoretical models, which are in good agreement with experimental results and comparable to the other theoretical  model calculations (M-M, RMF). The $\alpha$-decay calculations are performed for the two decay chains of $^{293,294}117$ nuclei keeping the parents and daughter in spherical nuclear shape and transtion of alpha decays are considered from ground state of the parents to the ground state of daughter nuclei. A maximum half-life at $^{282}111$ ($Z=111, N=171$) in ICM($Q^{Expt.}$) calculation is observed indicates the shell effects of the parent nucleus, near to the proposed shell closure (N=172) by other RMF model calculations.

\newpage

\begin{table*}
\caption{Comparison of calculated Isospin Cluster Model(ICM) $\alpha$-decay half-lives with experimental results, Macroscopic-Microscopic (M-M) \cite{sobi10} and with RMF model calculations (using NL3 \cite{patra11} and FSU Gold forces\cite{roy11}parameters). $Q^{ICM}$ is calculated using binding energies from BWMF \cite{samanta02}.}
\begin{tabular}{c   c  c  c  c  c  c c  c  l}\hline\hline
 Parent &$Q_{\alpha}^{Exp}$\cite{ogane10}&$Q_{\alpha}^{MM}$\cite{sobi10} &$Q_{\alpha}^{ICM}$\cite{samanta02} &Exp.\cite{ogane10} &MM\cite{sobi10} &ICM &ICM ($Q_{\alpha}^{Exp.}$ )&NL3\cite{patra11}&FSU\cite{roy11}   \\
 $^{A}Z$&(MeV) &(MeV) &(MeV) &$Log_{10}T_{1/2}(s)$ &$Log_{10}T_{1/2}(s)$ &$Log_{10}T_{1/2}(s)$ &$Log_{10}T_{1/2}(s)$&$Log_{10}T_{1/2}(s)$ &$Log_{10}T_{1/2}(s)$  \\\hline

$^{293}117$&11.183&11.42&11.699&-1.837&-2.159&-0.797&0.059&-2.71&-2.54 \\
$^{289}115$&10.45&10.63&11.100&-0.654&-0.817&-0.149&1.041&-0.96&-1.346 \\
$^{285}113$&9.88&10.10&10.495&0.738&-0.079&0.622&1.87&0.60&0.367 \\ \\
$^{294}117$&10.97&11.15&11.56&-1.109&-1.506&-0.402&0.269&-0.906&0.036 \\
$^{290}115$&10.10&10.37&10.95&1.797&-0.159&0.147&1.817&-2.325&0.226 \\
$^{286}113$&9.77&9.70&10.34&1.293&-1.955&0.941&2.139&1.055&0.4639 \\
$^{282}111$&9.13&9.57&9.73&-0.289&0.749&1.805&3.202&2.027&1.686 \\
$^{278}109$&9.67&9.27&9.11&0.882&0.955&2.821&1.55&2.316&-0.0682 \\
$^{274}107$&8.93&8.55&8.47&1.732&2.488&4.06&2.89&3.913&0.893 \\

\\\hline\hline

\end{tabular}

\end{table*}

\end{document}